\documentclass[aps,twocolumn,floatfix,superscriptaddress]{revtex4-2}

\usepackage{bm,bbm,amsmath,amssymb,amsbsy,amsthm,graphicx,subfigure,color,txfonts,multirow,enumitem}
\usepackage[colorlinks]{hyperref}
\hypersetup{
	bookmarksnumbered,
	pdfstartview={FitH},
	citecolor={blue},
	linkcolor={red},
	urlcolor={black},
	pdfpagemode={UseOutlines}
}
\usepackage{cleveref}
\usepackage{amsfonts}
\usepackage{float}
\usepackage{tikz}
\usepackage[singlelinecheck=false,justification=raggedright]{caption} 
\usepackage{tabularx}

\theoremstyle{definition}

\allowdisplaybreaks

\begin{document}
\title{Evaluating Pauli errors on cluster states by weighted distances}

\author{Choong Pak Shen}
\affiliation{UCSI College, Lot 39890, Jalan Cerdas Taman Connaught Cheras, 56000 Kuala Lumpur, Malaysia}
\affiliation{Institute for Mathematical Research, Universiti Putra Malaysia, 43400 Serdang, Selangor, Malaysia}
\author{Davide Girolami}
\affiliation{Politecnico di Torino, Corso Duca Degli Abruzzi, 24, 10129 Torino, Italy}
\date{\today}

\begin{abstract}
We address the problem of evaluating the difference between quantum states before and after being affected by errors encoded in unitary transformations. Standard distance functions, e.g., the Bures length, are not fully adequate for such a task. Weighted distances are instead appropriate information measures to quantify distinguishability of multipartite states. 
Here, we employ the previously introduced weighted Bures length and the newly defined weighted Hilbert-Schmidt distance to quantify how much single-qubit Pauli errors alter cluster states. We find that different errors of the same dimension change cluster states in a different way, i.e., their detectability is in general different. Indeed, they transform an ideal cluster state into a state whose weighted distance from the input depends on the specific chosen Pauli rotation, as well as the position of the affected qubit in the graph related to the  state. As these features are undetected by using standard distances, the study  proves the usefulness of weighted distances to monitor key but elusive properties of many-body quantum systems.
\end{abstract}

\maketitle

\section*{Introduction}

Quantifying the difference between quantum states is a critical task in both theoretical and experimental quantum information processing. Indeed, several quantifiers of state difference have been invented \cite{book}. An important class of them is represented by geometric distance functions, which are built to compare the physical properties of different configurations in terms of their distance in the abstract space of all possible quantum states of a system \cite{spehner,revfid}. 

Despite their wide applicability and obvious importance, such ``standard" distances are not fully adequate to investigate quantum states of many particles, as they do not take into account their dimension. For such a reason, the concept of {\it weighted} distances has been introduced \cite{Girolami2021}. These quantities satisfy a set of desirable mathematical properties, just as standard distances, while factoring in the dimension of the system under scrutiny and, at the same time, the size of the apparatuses that implement appropriate discriminating measurements.

In particular, the weighted Bures length has a compelling operational interpretation: It quantifies the difficulty in distinguishing two quantum states by the most informative measurements. 
 Subsequent studies have found that the weighted Bures length between the input and output states of a computation quantifies  the complexity of the process \cite{w1}, and signals chaotic behaviour in quantum dynamics \cite{w2,w3}. \\
 

In this paper, we expand the use of weighted distances, by applying them to evaluate the difficulty of detecting errors, i.e., uncontrollable, unwanted changes of a state. Specifically, we calculate the weighted Bures length, and a newly introduced weighted Hilbert-Schmidt distance, between cluster states of three, four, and five qubits, and the same states after they have undergone unplanned single-qubit unitary rotations (``Pauli errors''). Cluster states are an important class of states that are entangled even when information about some system components is lost. That is, they are ``robust", as they display genuine multipartite entanglement \cite{cluster3,cluster5,cluster6}. Also, they are the resource to perform one-way quantum computations \cite{cluster1,cluster2,cluster4}. 

Our results show that there exists a hierarchy between unitary errors in terms of their effects on an initial, ideal cluster state. Specifically, input cluster states are sent by single-qubit Pauli errors into output states whose weighted distance from the input depends on the chosen Pauli error. That is, our ability to detect when a cluster state has been  corrupted  depends on the specific error. \\
Moreover, we observe another surprising feature. As the number of qubits increases, phase flips (Pauli-$Z$ errors) that affect   ``central'' qubits of one-dimensional cluster states send the system into a state that is closer to the input state in comparison with phase flips affecting ``peripheral'' qubits. Centrally located errors are therefore more difficult to detect, somehow hidden in the structure of the system.\\
Remarkably, both results are independent of the employed weighted distance, as they describe how errors change subsystems of different size within systems prepared in cluster states. Also, neither phenomena can be observed by using any standard distances.

It is important to note that state transformation and state discrimination  are ``quite different tasks'' \cite{woot2}. Indeed, every single unitary error can be corrected by reapplying the very same unitary.  More generally, quantifiers of the complexity of an input/output transformation (how difficult it is to transform quantum states) and input/output distinguishability (how difficult it is to distinguish quantum states) are in general different kinds of information measures. Yet, the weighted Bures length, which reliably evaluates our ability to discriminate between two states  via measurements, is also a lower bound to the experimental cost, in terms of physical resources, to transform the states into each other \cite{Girolami2021}. \\

The paper is organized as follows. In Section \ref{sec1}, we review the notion of weighted distances and their advantages with respect to standard geometric quantifiers of state difference. First, we recall the definition of Bures length introduced in \cite{Girolami2021}. Then, we introduce the computationally friendly weighted Hilbert-Schmidt distance and prove some properties that inherits from the related standard distance. In Section \ref{sec2}, we present a case study: We evaluate the difference between cluster states plagued by unitary errors and their  textbook form in terms of their weighted distances. In the Conclusion, we draw our final comments and suggest further lines of research.

\section{From standard distances to weighted distances} \label{sec1}

The difference between quantum states, e.g., between a target configuration and what one actually implements in an experiment, is customarily quantified by means of state fidelity. The fidelity between two pure states is indeed equivalent to the state overlap:
\begin{align}
F(\left| \psi \right\rangle,\, \left| \phi \right\rangle) = \left| \langle \psi | \phi \rangle \right|^2.
\end{align}
Remarkably, the overlap has a geometric interpretation, being related to the Fubini-Study distance $\cos^{-1}\sqrt{F(\left| \psi \right\rangle,\, \left| \phi \right\rangle)}$ \cite{book,wootters,fubini,study}. Generalizing the fidelity to mixed states \cite{Uhlmann1976,jozsa,fuchs,Gilchrist2005}, one has
\begin{align}
F(\rho,\, \sigma) = \left(\text{Tr} \sqrt{\sqrt{\rho} \sigma \sqrt{\rho}} \right)^2. \label{fidelity}
\end{align}
Among many distance functions for mixed states, the Bures length between two density matrices $\rho,\sigma$ is the natural generalization of the Fubini-Study distance:
\begin{align}
B(\rho,\, \sigma) = \text{cos}^{-1} \sqrt{F(\rho,\, \sigma)}.
\end{align}
State fidelity and Bures length are ubiquitous in quantum information theory. They are well motivated quantities, as they provably quantify the difficulty in discriminating quantum states by means of a single measurement on the global system. Further, they are the most popular quantifiers of how close an engineered quantum state, say $\rho_N$, is to the desired target $\sigma_N$.

Yet, they are not fully satisfactory metrics when one wants to compare states of multipartite quantum systems. Consider for example the following quantum states: 
\begin{align*}
 \left| 0 \right\rangle^{\otimes N}, 
\left| 0 \right\rangle^{\otimes N-k} \left| 1 \right\rangle^{\otimes k}.
\end{align*}
If we evaluate their difference by means of their Bures length, they are maximally far for any value  of $k$. Yet, the larger is $k$, the easier it is to experimentally distinguish them, as there are more (local) measurements that enable to discriminate  between these two states. In general, overlap-based distances reach maximal value when two global states are orthogonal, no matter how close the marginal density matrices may be. 
Another relevant example is the comparison between a state of $N$ qubits that displays maximal entanglement, say a GHZ state, with a classically correlated mixed state:
\begin{align*}
 \left(| 0 \rangle^{\otimes N} + | 1\rangle^{\otimes N}\right)/\sqrt2,\,\,  \frac{1}{2} \left(| 0 \rangle\langle 0 |^{\otimes N} + | 1 \rangle\langle 1 |^{\otimes N}\right).
\end{align*}
The difficulty of experimentally discriminating these two states increases with $N$, as only global measurements over all qubits would output different data, while the Bures length between them is independent of the system size. \\

In order to overcome these issues, while retaining the explanatory power of distance functions, the concept of weighted distances was introduced \cite{Girolami2021}. They are generalizations of standard distances that take into account the size of both the system and the possible measurements one can perform on it.

We review the construction of the weighted Bures length between two preparations $\rho_N, \sigma_N$ of an $N$-particle system (see \cite{Girolami2021} for full details), while the argument applies to any distance function. Suppose cooperating agents independently measure on different subsystems $k_\alpha \leq N$ of two copies of the system, which are prepared in the states $\rho_N, \sigma_N$, respectively. Each agent evaluates the difference between two states of the assigned subsystem by the Bures length $B(\rho_{k_\alpha},\, \sigma_{k_\alpha})$. In particular, they perform the optimal measurement ${\cal M}^{k_\alpha}=\rho_{k_\alpha}^{-1/2}\sqrt{\rho_{k_\alpha}^{1/2}\sigma_{k_\alpha} \rho_{k_\alpha}^{1/2}}\rho_{k_\alpha}^{-1/2}$ to discriminate the marginal states of $k_\alpha$ particles. The setup defines a measurement partition $P_{k_\alpha}:=\left\{{\cal M}^{k_\alpha},\,\sum_a\,k_\alpha=N\right\}$.

Then, one builds a weighted sum of each agent contribution $B(\rho_{k_\alpha},\, \sigma_{k_\alpha})$, assigning a weight $1/k_\alpha$ to each of them, i.e., evaluating their importance as inversely proportional to the size of the related subsystem. The reason is the following: One can quantify the difficulty in realizing a measurement with the size of the measured subsystem $k_\alpha$. As a distance must quantify how easy it is to distinguish the two states, i.e., how easy it is to experimentally carry out the related measurement, a sound choice of weights is $1/k_\alpha$. Finally, one defines the weighted Bures length by optimizing the measurement strategy, which implies to maximize the weighted sum of local distances over all possible partitions: 
\begin{align}
D_B (\rho,\, \sigma) := \max\limits_{P_{k_\alpha}} \sum_a \frac{1}{k_\alpha} B\left(\rho_{k_\alpha},\, \sigma_{k_\alpha}\right). \label{weightedBureslength}
\end{align}

The weighted Bures length meets a set of desirable properties:
\begin{itemize}
\item $D_B\rho_N,\sigma_N) \geq 0$ (non-negativity)

\item $D_B(\rho_N,\sigma_N) = 0 \iff \rho_N = \sigma_N$ (faithfulness)

\item $D_B(\rho_N,\sigma_N) \geq D(\Lambda_1 (\rho_N),\Lambda_1 (\sigma_N)),\,\, \forall \Lambda_1$ (contractivity under single qubit operations $\Lambda_1$)

\item $D_B(\rho_N,\sigma_N) \leq D_B(\rho_N,\tau_N) + D_B(\tau_N, \sigma_N)$ (triangle inequality)
\end{itemize}
The weighted Bures length has also a second operational meaning that is not shared by the standard Bures length. It is shown to be the lower bound of experimental cost of the state transformation $\rho_N = U\sigma_N U^{\dagger}, U=e^{-i\,H\,t}$:
\begin{equation}
N\,E\,t \geq D_B(\rho_N,\sigma_N),
\end{equation}
in which $E = (\lambda_{max} - \lambda_{min})/2$, being $\lambda_{max(min)}$ the largest (smallest) eigenvalue of the Hamiltonian. Hence, the weighted Bures length lower bounds the cost of transforming $\rho_N$ into $\sigma_N$ (and vice versa), as quantified by the product of the size of the system $N$, the energy parameter $E$, and the available time $t$ \cite{Girolami2021}.\\

While the weighted Bures length enjoys important properties, one can build a parent weighted distance for each standard one. Here, we define the weighted Hilbert-Schmidt distance:
\begin{align}\label{HS}
D_{HS}(\rho_N,\sigma_N):= \max\limits_{P_{k_\alpha}} \sum_a \frac{1}{k_\alpha} d_{HS}(\rho_{k_\alpha},\sigma_{k_\alpha}),
\end{align}
where $d_{HS}$ is the (square root of the) standard Hilbert-Schmidt distance, defined as
\begin{align}
d_{HS}(\rho_N,\sigma_N):=  \sqrt{\text{Tr}\left(\rho_N^2 +\sigma_N^2-2\,\rho_N \sigma_N\right)}.
\end{align}
Note that the square root is here crucial. We will  run a numerical comparison with the weighted Bures length, so both quantities need to be functions of the state eigenvalues elevated to the same powers.
 The weighted Hilbert-Schmidt distance inherits the following properties:
\begin{itemize}
\item $D_{HS}(\rho_N,\sigma_N) \geq 0$ (non-negativity)\\
{\it Proof}: A weighted distance is the sum of standard distances with positive weights, so the property is satisfied.

\item $D_{HS}(\rho_N,\sigma_N) = 0 \iff \rho_N = \sigma_N$ (faithfulness)\\
{\it Proof}: Since all the weights are positive, the weighted sum for an arbitrary partition $P_{k_\alpha}$ is non-zero if and only if there is at least a non-zero term $d_{HS}(\rho_{k_{\alpha}},\sigma_{k_\alpha})$. Hence, the claim is proven.

\item $D_{HS}(\rho_N,\sigma_N)\leq  D_{HS}(\rho_N,\tau_N) + D_{HS}(\tau_N, \sigma_N)$ (triangle inequality)\\
{\it Proof}: For an arbitrary term of the weighted sum related to an arbitrary partition, one has $\frac{1}{k_\alpha} d_{HS}\left(\rho_{k_\alpha},\sigma_{k_\alpha}\right) \leq \frac{1}{k_\alpha} \left\{d_{HS}\left(\rho_{k_\alpha},\tau_{k_\alpha}\right) + d_{HS}\left(\tau_{k_\alpha},\sigma_{k_\alpha}\right)\right\}$. Calling $\hat k_\alpha$ the weights related to the partition that maximizes the weighted sum of the terms $d_{HS}\left(\rho_{k_\alpha},\sigma_{k_\alpha}\right)$, one has
\begin{align*}
D_{HS}(\rho_N,\sigma_N)
& = \sum_{\alpha}\frac{1}{\hat k_\alpha}d_{HS}\left(\rho_{\hat k_\alpha},\sigma_{\hat k_\alpha}\right) \\
& \leq \sum_{\alpha}\frac{\left\{d_{HS}\left(\rho_{\hat k_\alpha},\tau_{\hat k_\alpha}\right) + d_{HS}\left(\tau_{\hat k_\alpha},\sigma_{\hat k_\alpha}\right)\right\}}{\hat k_\alpha}\\
& \leq D_{HS}(\rho_N,\tau_N) + D_{HS}(\tau_N,\sigma_N).
\end{align*}
\end{itemize}

Note that the weighted Hilbert-Schmidt distance is not monotonically decreasing under local operations, as the related standard  distance can increase under noisy channels \cite{spehner,schirmer}. Yet, since the Hilbert-Schmidt distance is a function of purities of the states and their overlaps, it is manifestly easy to compute. Also, it is experimentally friendly. Indeed, state purities and overlaps can be quantified without full state tomography and full spectrum reconstruction \cite{Zhang2017,det1,det2,smith,det3,det4,det5}.\\

In the next section, we will run a comparison of these two quantities to evaluate the difference between an ideal state and its corrupted version. We will observe that the same subset of single-qubit Pauli errors  generate states closer to an input cluster state according to both the weighted Bures length and the weighted Hilbert-Schmidt distance, even if the latter is not in general contractive under quantum channels.

\section{Case Study: Evaluation of unitary errors in cluster states by weighted distances}\label{sec2}
{
\captionsetup[table]{name=Figure}
\begin{table}[h]\
\begin{tabular}{cc}
\begin{tikzpicture}[main_node/.style={circle,fill=blue!20,draw,minimum size=1em,inner sep=3pt}]
\node[main_node] (1) at (0,0) {1};
\node[main_node] (2) at (1,0) {2};
\node[main_node] (3) at (2,0) {3};
\draw (1) -- (2) -- (3);
\end{tikzpicture}
&
\begin{tikzpicture}[main_node/.style={circle,fill=blue!20,draw,minimum size=1em,inner sep=3pt}]
\node[main_node] (1) at (0,0) {1};
\node[main_node] (2) at (1,0) {2};
\node[main_node] (3) at (2,0) {3};
\node[main_node] (4) at (3,0) {4};
\draw (1) -- (2) -- (3) -- (4);
\end{tikzpicture}
\\
(a) & (b) \\
\begin{tikzpicture}[main_node/.style={circle,fill=blue!20,draw,minimum size=1em,inner sep=3pt}]
\node[main_node] (1) at (0,0) {1};
\node[main_node] (2) at (1,0) {2};
\node[main_node] (3) at (2,0) {3};
\node[main_node] (4) at (3,0) {4};
\node[main_node] (5) at (4,0) {5};
\draw (1) -- (2) -- (3) -- (4) -- (5);
\end{tikzpicture}
&
\begin{tikzpicture}[main_node/.style={circle,fill=blue!20,draw,minimum size=1em,inner sep=3pt}]
\node[main_node] (1) at (0,0) {1};
\node[main_node] (2) at (1,0) {2};
\node[main_node] (3) at (1,-1) {3};
\node[main_node] (4) at (0,-1) {4};
\draw (1) -- (2) -- (3) -- (4) -- (1);
\end{tikzpicture}
\\
(c) & (d)
\end{tabular}
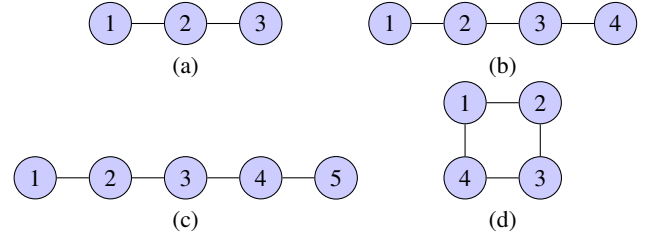
\caption{Graphs (a)-(d) depict the four different configurations of cluster states that are here under scrutiny.} \label{clusterstates}
\end{table}
}

\subsection{Cluster states and Pauli errors}
Cluster states are a subset of graph states which are simultaneous eigenstates with eigenvalue one of commuting Pauli operators. They can be employed to perform one-way (also called ``measurement-based") quantum computations \cite{cluster1,cluster2,cluster4}. By using the stabilizer formalism, the density matrix of $N$-qubit cluster states can be written in a compact form \cite{cluster5,Tiurev2022}:
\begin{align}
\rho = \prod_{i=1}^N \frac{I_i + g_i}{2}, \label{clusterstatesequation}
\end{align}
where $I_i$ is the bidimensional identity matrix and $g_i = X_i \bigotimes_{j \in N(i)} Z_j$ are the stabilizing operators. Here, $X_i$ and $Z_j$ are Pauli $X$ and $Z$ matrices, respectively, while $N(i)$ denotes the neighbourhood of $i$-th qubit (consisting of all the qubits that are adjacent to the $i$-th qubit in the graphical representation of the state). Each cluster state represents a specific correlation structure of qubits placed at vertices of a lattice.

The stabilizer formalism of graph states, including cluster states, allows us to write their reduced density matrices in terms of their stabilizers \cite{cluster3,cluster6}. For the sake of clarity, we provide the Reader with the explicit expressions of the global and marginal density matrices of the cluster states that we selected for this study (identity matrices are omitted when possible, e.g., $X_1X_2 \equiv X_1X_2I_{345}$), which are also depicted in Figure \ref{clusterstates}.

{\it Three-qubit one-dimensional cluster states:}
\begin{align*}
\rho_{123} & = \frac{1}{8} (I_{123} + X_1 Z_2) (I_{123} + Z_1 X_2 Z_3) (I_{123} + Z_2 X_3) \nonumber \\
& = \frac{1}{8} (I_{123} + Z_2 X_3 + X_1 Z_2 + X_1 X_3 \nonumber \\
& \quad + Z_1 X_2 Z_3 + Z_1 Y_2 Y_3+ Y_1 Y_2 Z_3 - Y_1 X_2 Y_3), \\
\rho_{12} & = \frac{1}{4} (I_{12} + X_1 Z_2), \\
\rho_{13} & = \frac{1}{4} (I_{13} + X_1 X_3), \\
\rho_{23} & = \frac{1}{4} (I_{23} + Z_2 X_3).
\end{align*}
{\it Four-qubit one-dimensional cluster states:}
\begin{align*}
\rho_{1234}^{1D} & = \frac{1}{16} (I_{1234} + X_1 Z_2) (I_{1234} + Z_1 X_2 Z_3) \nonumber \\
& \quad (I_{1234} + Z_2 X_3 Z_4)(I_{1234}  + Z_3 X_4) \nonumber \\
& = \frac{1}{16} (I_{1234} + Z_3 X_4 + X_1 Z_2 + Z_2 X_3 Z_4 + Z_2 Y_3 Y_4 \nonumber \\
& \quad + Z_1 X_2 Z_3 + Z_1 X_2 X_4 + X_1 X_3 Z_4 + X_1 Y_3 Y_4 \nonumber \\
& \quad + Y_1 Y_2 Z_3 + Y_1 Y_2 X_4 + Z_1 Y_2 Y_3 Z_4 - Z_1 Y_2 X_3 Y_4 \nonumber \\
& \quad + X_1 Z_2 Z_3 X_4 - Y_1 X_2 Y_3 Z_4 + Y_1 X_2 X_3 Y_4), \\
\rho_{123}^{1D} & = \frac{1}{8} (I_{123} + X_1 Z_2) (I_{123} + Z_1 X_2 Z_3), \\
\rho_{124}^{1D} & = \frac{1}{8} (I_{124} + X_1 Z_2) (I_{124} + Z_1 X_2 X_4), \\
\rho_{134}^{1D} & = \frac{1}{8} (I_{134} + Z_3 X_4) (I_{134} + X_1 X_3 Z_4), \\
\rho_{234}^{1D} & = \frac{1}{8} (I_{234} + Z_2 X_3 Z_4) (I_{234} + Z_3 X_4), \\
\rho_{12}^{1D} & = \frac{1}{4} (I_{12}+ X_1 Z_2), \\
\rho_{34}^{1D} & = \frac{1}{4} (I_{34} + Z_3 X_4), \\
\rho_{13}^{1D} & = \rho_{14}^{1D} = \rho_{23}^{1D} = \rho_{24}^{1D} = \frac{1}{4} I_{ij,\,\, i=1,2, j=3,4}.
\end{align*}
{\it Four-qubit two-dimensional cluster states:}
\begin{align*}
\rho_{1234}^{2D} & = \frac{1}{16} (I_{1234} + Z_4 X_1 Z_2) (I_{1234}  + Z_1 X_2 Z_3) \nonumber\\
& \quad(I_{1234} + Z_2 X_3 Z_4)(I_{1234}  + Z_1 X_4 Z_3) \nonumber \\
& = \frac{1}{16} (I_{1234} + X_1 X_3 + X_2 X_4 + Z_1 X_4 Z_3 + Z_2 X_3 Z_4 \nonumber \\
& \quad - Y_2 X_3 Z_4 + X_1 Z_2 Z_4 - Y_1 Y_3 X_4 - X_1 Y_2 Y_4 \nonumber \\
& \quad - Y_1 X_2 Y_3 + Z_1 X_2 Z_3 + Z_1 Z_2 Y_3 Y_4 + Z_1 Y_2 Y_3 Z_4 \nonumber \\
& \quad + Y_1 Z_2 Z_3 Y_4 + Y_1 Y_2 Z_3 Z_4 + X_1 X_2 X_3 X_4), \\
\rho_{123}^{2D} & = \frac{1}{8} (I_{123} + Z_1 X_2 Z_3) (I_{123} + X_1 X_3), \\
\rho_{124}^{2D} & = \frac{1}{8} (I_{124} + Z_4 X_1 Z_2) (I_{124} + X_2 X_4), \\
\rho_{134}^{2D} & = \frac{1}{8} (I_{134} + Z_1 X_4 Z_3) (I_{134} + X_1 X_3), \\
\rho_{234}^{2D} & = \frac{1}{8} (I_{234} + Z_2 X_3 Z_4) (I_{234} + X_2 X_4), \\
\rho_{13}^{2D} & = \frac{1}{4} (I_{13} + X_1 X_3), \\
\rho_{24}^{2D} & = \frac{1}{4} (I_{24} + X_2 X_4), \\
\rho_{12}^{2D} & = \rho_{14}^{2D} = \rho_{23}^{2D} = \rho_{34}^{2D} = \frac{1}{4} I_{ij,\,\, ij = \{ 12, 14, 23, 34 \}}.
\end{align*}
{\it Five-qubit one-dimensional cluster states:}
\begin{align*}
\rho_{12345} & = \frac{1}{32} (I_{12345} + X_1 Z_2) (I_{12345} + Z_1 X_2 Z_3) \nonumber \\
& \quad (I_{12345} + Z_2 X_3 Z_4)(I_{12345} + Z_3 X_4 Z_5) \nonumber \\
& \quad (I_{12345} + Z_4 X_5) \nonumber \\
& = \frac{1}{32} (I_{12345} + Z_4 X_5 + X_1 Z_2 + Z_3 X_4 Z_5 + Z_3 Y_4 Y_5 \nonumber \\
& \quad + Z_2 X_3 Z_4 + Z_2 X_3 X_5 + Z_1 X_2 Z_3 + X_1 X_3 Z_4 \nonumber \\
& \quad + X_1 X_3 X_5 + Y_1 Y_2 Z_3 + Z_2 Y_3 Y_4 Z_5 \nonumber \\
& \quad - Z_2 Y_3 X_4 Y_5 + Z_1 X_2 X_4 Z_5 + Z_1 X_2 Y_4 Y_5 \nonumber \\
& \quad + Z_1 Y_2 Y_3 Z_4 + Z_1 Y_2 Y_3 X_5 + X_1 Z_2 Z_4 X_5 \nonumber \\
& \quad + X_1 Y_3 Y_4 Z_5 - X_1 Y_3 X_4 Y_5 + Y_1 Y_2 X_4 Z_5 \nonumber \\
& \quad + Y_1 Y_2 Y_4 Y_5 - Y_1 X_2 Y_3 Z_4 - Y_1 X_2 Y_3 X_5 \nonumber \\
& \quad + Z_1 X_2 Z_3 Z_4 X_5 - Z_1 Y_2 X_3 Y_4 Z_5 + Z_1 Y_2 X_3 X_4 Y_5 \nonumber \\
& \quad + X_1 Z_2 Z_3 X_4 Z_5 + X_1 Z_2 Z_3 Y_4 Y_5 + Y_1 Y_2 Z_3 Z_4 X_5 \nonumber \\
& \quad + Y_1 X_2 X_3 Y_4 Z_5 - Y_1 X_2 X_3 X_4 Y_5), \\
\rho_{1234} & = \frac{1}{16} (I_{1234} + X_1 Z_2) (I_{1234} + Z_1 X_2 Z_3) (I_{1234} + Z_2 X_3 Z_4), \\
\rho_{1235} & = \frac{1}{16} (I_{1235} + X_1 Z_2) (I_{1235} + Z_1 X_2 Z_3) (I_{1235}+ Z_2 X_3 X_5), \\
\rho_{1245} & = \frac{1}{16} (I_{1245} + X_1 Z_2) (I_{1245} + Z_4 X_5)\\ &\quad (I_{1245}  + Z_1 X_2 X_4 Z_5), \\
\rho_{1345} & = \frac{1}{16} (I_{1345} + Z_3 X_4 Z_5) (I_{1345} + Z_4 X_5) \\ & \quad (I_{1345} + X_1 X_3 Z_4), \\
\rho_{2345} & = \frac{1}{16} (I_{2345} + Z_2 X_3 Z_4) (I_{2345} + Z_3 X_4 Z_5)\\ & \quad (I_{2345} + Z_4 X_5), \\
\rho_{123} & = \frac{1}{8} (I_{123} + X_1 Z_2) (I_{123} + Z_1 X_2 Z_3), \\
\rho_{124} & = \rho_{125} = \frac{1}{8} (I_{12i,\, i=4,5} + X_1 Z_2), \\
\rho_{134} & = \frac{1}{8} (I_{134} + X_1 X_3 Z_4), \\
\rho_{135} & = \frac{1}{8} (I_{135} + X_1 X_3 X_5), \\
\rho_{145} & = \rho_{245} = \frac{1}{8} (I_{i45,\,i=1,2} + Z_4 X_5), \\
\rho_{234} & = \frac{1}{8} (I_{234} + Z_2 X_3 Z_4), \\
\rho_{235} & = \frac{1}{8} (I_{235} + Z_2 X_3 X_5), \\
\rho_{345} & = \frac{1}{8} (I_{345} + Z_3 X_4 Z_5) (I_{345} + Z_4 X_5), \\
\rho_{12} & = \frac{1}{4} (I_{12} + X_1 Z_2), \\
\rho_{45} & = \frac{1}{4} (I_{45} + Z_4 X_5), \\
\rho_{13} & = \rho_{14} = \rho_{15} = \rho_{23} = \rho_{24} = \rho_{25} = \rho_{34} \\
& = \rho_{35} = \frac{1}{4} I_{ij,\,i=1,2,3,j=3,4,5,\,i \neq j}.
\end{align*}

We will now evaluate the detectability of  Pauli errors when they are perturbing these  cluster states. 
Any single-qubit Pauli errors will transform cluster states into orthogonal states. Thus, standard distance functions, when being computed between the ideal cluster state and its corrupted version, will always reach the maximal value, regardless of how many and what kind of single-qubit Pauli errors are present in the cluster states. In other words, overlap-based distances are not fully adequate to discriminate how much different errors affect a quantum computation. 
Since the preparation of cluster states only requires controlled-$Z$ gates to act between the nearest-neighbouring qubits in the $\left| + \right\rangle$ basis, the most common experimental errors on cluster states are single-qubit unwanted rotations (bit/phase flips) of a specific system component or of the neighbouring  qubits \cite{cluster1,cluster2,cluster3,cluster4,cluster5,cluster6,Tiurev2022}. Hence, it is important to be able to properly quantify the influence of single-qubit Pauli errors on cluster states, as  they are arguably the most relevant ones for practical purposes. 

\subsection{Evaluation of Pauli errors by the weighted Bures length}

Unlike standard distance functions, weighted distances take into the consideration the statistical distinguishability of marginal as well global density matrices, in order to assess the difference between two quantum states. The construction is particularly appealing when we want to compare states that are invariant under (all possible, or at least some) subsystem permutations. In such a case, the maximization of the weighted sum over all  system partitions in Eq. (\ref{weightedBureslength}) is easily computable.\\

First, we employ the weighted Bures length to quantify how cluster states are affected by Pauli errors. We consider four different cluster states: Three-, four- and five qubit one-dimensional cluster states, and four-qubit two-dimensional cluster states, as shown in Figure \ref{clusterstates}. We show how different single-qubit rotations change such states, by evaluating the weighted Bures length between the cluster states before and after the error is applied.\\
The results are reported in Table \ref{weightedbureslengthpaulierrors}. They show that, surprisingly, different single-qubit rotations generate states that are placed at different distances from the original cluster states in the manifold of quantum states. Hence, the detectability of such errors is different.
For instance, consider the $Z_2$ error on four-qubit one- and two-dimensional cluster states. The possible elements of a four-qubit partition $P_{k_\alpha}$ are $\{ \{ 1 \}\mathrel{,} \{ 2 \}\mathrel{,} \{ 3 \}\mathrel{,} \{ 4 \}\mathrel{,} \{ 12 \}\mathrel{,} \{ 13 \}\mathrel{,} \{ 14 \}\mathrel{,} \{ 23 \}\mathrel{,} \{ 24 \}\mathrel{,} \{ 34 \}\mathrel{,} \{ 123 \}\mathrel{,} \{ 124 \}\mathrel{,} \{ 134 \}\mathrel{,} \{ 234\}\mathrel{,} \{ 1234 \}\}$. Since the one-body reduced density matrices of four-qubit cluster states are maximally mixed, local unitary operations will not be detectable by comparing their one-body reduced density matrices: $B(\rho_i,\sigma_i)=0,\,i=1,2,3,4$, in which $\rho_i, \sigma_i$ are the cluster state input and output single-qubit marginal density matrices, respectively. However,  the $Z_2$ error changes the  two-body  density matrix  of four-qubit two-dimensional cluster states  $\rho_{24}^{2D}$.

On the other hand, the $Z_2$ error cannot be detected by any of the two-body marginals of four-qubit one-dimensional cluster states, so in this case the output state is closer to the ideal cluster state. Instead, it changes the three-body marginals like $\rho_{123}^{1D} = \frac{1}{8} (I_{123} + X_1 Z_2) (I_{123} + Z_1 X_2 Z_3)$. Therefore, the effect of the $Z_2$ gate on four-qubit one-dimensional cluster states is different than on four-qubit two-dimensional cluster states, as captured by different values of weighted Bures length. The reason is that the system partition that maximizes the weighted sum in Eq.(\ref{weightedBureslength}) is different.


{
\captionsetup[table]{name=Table}
\setcounter{table}{0}
\begin{table}[h]
\centering
\begin{tabularx}{\columnwidth}{|X|X|X|}
\hline
$\qquad${\bf Cluster states} &  $\qquad \mathbf{D_{\mathbf B} =  \pi/6}$ & $\qquad\mathbf{D_{\mathbf B} =  \pi/4}$ \\
\hline
Three-qubit  \newline  one-dimensional & $X_1,\mathrel{\,} X_3,\mathrel{\,}Z_2$ & $Z_1,\mathrel{\,} Z_3,\mathrel{\,} X_2,\mathrel{\,} Y_1,\mathrel{\,} Y_2,\mathrel{\,}Y_3$  \\
\hline
Four-qubit \newline  one-dimensional & $X_1,\mathrel{\,} X_4,\mathrel{\,} Z_2,\mathrel{\,} Z_3$ & $Z_1,\mathrel{\,} Z_4,\mathrel{\,} X_2,\mathrel{\,} X_3,\mathrel{\,} Y_1,\mathrel{\,} Y_2,\mathrel{\,} Y_3,\mathrel{\,} Y_4$ \\
\hline
Four-qubit \newline  two-dimensional & $X_1,\mathrel{\,} X_2,\mathrel{\,} X_3,\mathrel{\,} X_4$ & $Z_1,\mathrel{\,} Z_2,\mathrel{\,} Z_3,\mathrel{\,} Z_4,\mathrel{\,} Y_1,\mathrel{\,} Y_2,\mathrel{\,} Y_3,\mathrel{\,} Y_4$ \\
\hline
Five-qubit \newline  one-dimensional &  $X_1,\mathrel{\,} X_3,\mathrel{\,} X_5,\mathrel{\,} Z_2,\mathrel{\,} Z_3,\mathrel{\,} Z_4,\mathrel{\,} Y_3$ & $Z_1,\mathrel{\,} Z_5,\mathrel{\,} X_2,\mathrel{\,} X_4,\mathrel{\,} Y_1,\mathrel{\,} Y_2,\mathrel{\,} Y_4,\mathrel{\,} Y_5$ \\
\hline
\end{tabularx}
\caption{Weighted Bures length between  cluster states before and after they have been affected by single-qubit Pauli errors. In the first column, we list the cluster states under scrutiny. In the second and third columns, for each cluster state, we identify the Pauli errors that generate output states whose weighted Bures length to the initial state are $\pi/6$ and $\pi/4$, respectively. Note that the standard Bures length reaches the maximal value $\pi/2$ in all these cases.} \label{weightedbureslengthpaulierrors}
\end{table}
}

%
%

Also, we observe that, for one dimensional states, the subset of single-qubit Pauli-$Z$ errors with a smaller weighted distance measure seems determined by the geometry of cluster states. Specifically, unwanted phase flips of ``central'' qubits ($Z_2$ for  three-qubit states, $Z_2,Z_3$ for four-qubit states, one dimensional states, and $Z_2,Z_3,Z_4$ for five-qubit states) are more difficult to detect than peripheral ones. 

\subsection{Evaluation of Pauli errors by weighted Hilbert-Schmidt distance}
{
\captionsetup[table]{name=Table}
\begin{table}[h]
\begin{tabularx}{\columnwidth}{|X|X|X|}
\hline
$\qquad${\bf Cluster states} & $\qquad\mathbf{D_{\mathbf HS} =\sqrt2/3}$ & $\qquad\mathbf{D_{\mathbf HS} = 1/2 }$ \\
\hline
Three-qubit \newline one-dimensional & $X_1,\mathrel{\,} X_3,\mathrel{\,}Z_2$ & $Z_1,\mathrel{\,} Z_3,\mathrel{\,} X_2,\mathrel{\,} Y_1,\mathrel{\,} Y_2,\mathrel{\,}Y_3$ \\
\hline
$\qquad${\bf Cluster states} & $\qquad\mathbf{D_{\mathbf HS} =\sqrt{2}/4}$ & $\qquad\mathbf{D_{\mathbf HS} = 1/2 }$ \\
\hline
Four-qubit \newline  one-dimensional & $X_1,\mathrel{\,} X_4,\mathrel{\,} Z_2,\mathrel{\,} Z_3$ & $Z_1,\mathrel{\,} Z_4,\mathrel{\,} X_2,\mathrel{\,} X_3,\mathrel{\,} Y_1,\mathrel{\,} Y_2,\mathrel{\,} Y_3,\mathrel{\,} Y_4$ \\
\hline
$\qquad${\bf Cluster states} & $\qquad\mathbf{D_{\mathbf HS} =\sqrt2/4}$ & $\qquad\mathbf{D_{\mathbf HS} = 1/2 }$ \\
\hline
Four-qubit \newline  two-dimensional & $X_1,\mathrel{\,} X_2,\mathrel{\,} X_3,\mathrel{\,} X_4$ & $Z_1,\mathrel{\,} Z_2,\mathrel{\,} Z_3,\mathrel{\,} Z_4,\mathrel{\,} Y_1,\mathrel{\,} Y_2,\mathrel{\,} Y_3,\mathrel{\,} Y_4$ \\
\hline
$\qquad${\bf Cluster states} & $\qquad\mathbf{D_{\mathbf HS} =1/3}$ & $\,\qquad\mathbf{D_{\mathbf HS} = 1/2 }$ \\
\hline
Five-qubit \newline  one-dimensional & $X_1,\mathrel{\,} X_3,\mathrel{\,} X_5,\mathrel{\,} Z_2,\mathrel{\,} Z_3,\mathrel{\,} Z_4,\mathrel{\,} Y_3$ & $Z_1,\mathrel{\,} Z_5,\mathrel{\,} X_2,\mathrel{\,} X_4,\mathrel{\,} Y_1,\mathrel{\,} Y_2,\mathrel{\,} Y_4,\mathrel{\,} Y_5$ \\
\hline
\end{tabularx}
\caption{Weighted Hilbert-Schmidt distance between  cluster states before and after they have been affected by single-qubit Pauli errors. In the first column, we list the cluster states under scrutiny. In the second  and third columns, for each cluster state, we list the Pauli errors and the related weighted Hilbert-Schmidt distance between input and output states. Note that the standard Hilbert-Schmidt distance takes the maximal value $\sqrt2$ in all these cases.} \label{HSlengthpaulierrors}
\end{table}
}
We now repeat the calculation by employing the weighted Hilbert-Schmidt distance introduced in Eq.(\ref{HS}), which is in general simpler to evaluate than the weighted Bures length and other distances. Also, we want to verify whether unitaries of the same dimension still generate states at different distances from the input. That is, we investigate if  this surprising fact  depends on the employed distance function, rather than being a generic feature of quantum dynamics.\\
As we said before, single-qubit Pauli errors transform ideal cluster states into orthogonal states. Therefore, the input/output global state overlap is zero. This means that the weighted Hilbert-Schmidt distance of cluster states to their alterations depends only on the global and marginal purities and the overlaps between the state marginals.\\

The results are reported in Table \ref{HSlengthpaulierrors}. As it happened by using the weighted Bures length, different errors yield different values of the weighted Hilbert-Schmidt distance. Remarkably, we can establish a hierarchy among Pauli matrices in terms of how much they affect the cluster states, independently of the employed distance. For example, for three-qubit cluster states, the very same $X_1,\, X_3,\, Z_2$ errors that affect the least the initial state, as quantified by a smaller weighted Bures length, also output the closest state according to the weighted Hilbert-Schmidt distance $\left(\sqrt2/3\approx 0.47\right)$. For the rest of the Pauli errors, the weighted Hilbert-Schmidt distance  is $1/2$. (The optimal partitition is  made of a pair of two-body marginal states.)
 In the case of four-qubit one-dimensional cluster states, given the $X_1,\, X_4,\, Z_2,\, Z_3$ errors, the optimal partition is the full four-body one (only global measurements can distnguish between the input/output states): therefore, the weighted Hilbert-Schmidt distance is $\sqrt2/4 \approx 0.35$. Again, all the other errors generate further states. Similarly, for $X_1,\, X_2,\, X_3,\, X_4$ errors, the weighted Hilbert-Schmidt distance between inputs and outputs for four-qubit two-dimensional states is $\sqrt2/4$. The least affecting errors for the five-qubit case are instead $X_1,\, X_3,\, X_5,\, Z_2,\, Z_3,\, Z_4,\, Y_3$: the optimal measurement is a three-body one and the weighted Hilbert-Schmidt distance takes the value $1/3$. \\
Finally, by employing the weighted Hilbert-Schmidt distance, we observe again how the structure of one-dimensional states affects the detectability of Pauli-$Z$ errors. Central $Z$-errors are still the most difficult to spot via measurements. We conjecture that, considering one-dimensional cluster states of increasing size,  Pauli $Z$ errors on the inner qubits of the cluster states would still generate states that are closer to the inputs. Also, more complex  two-dimensional states than the one studied here may be also prone to hide centrally located errors.

\section*{Conclusion}
We evaluated how local unitary errors alter cluster states by means of the weighted Bures length and the newly defined weighted Hilbert-Schmidt distance. We showed that, while all single-qubit Pauli errors transform ideal cluster states into orthogonal erroneous states, the distinguishability of such corrupted states with respect to the input configuration can be different. Unitary errors of the same dimension produce states that have different weighted distance from the initial state, which physically means that the size of the optimal measurements that detect them depends on the specific error. 

Our results show how weighted distances can be more informative than standard distances  in evaluating important properties of many-body quantum systems. As potential follow-on projects, we anticipate that, just like standard distances, weighted distances can be split into genuine ``quantum" and ``classical" parts \cite{howdiff}. These more refined weighted quantum distances could be employed to evaluate how errors affect quantum resources rather than the whole quantum states. This is much needed to investigate properties of large-dimensional entangled states \cite{dicke}, and their resistance to various error sources. Other interesting avenues of investigation are extensions of our case study to other classes of multipartite quantum states with complex correlation structure, e.g., other types of graph states. Finally, our findings call for a rigorous proof that the structure of cluster states determines what Pauli errors are more difficult to detect, regardless of the employed weighted distance.

\section*{Acknowledgments}

C. P. S. is thankful to Kwek Leong Chuan and Konstantin Tiurev for private discussions. C. P. S. is sponsored by Erasmus+ EU Funds for the duration of study. D. G. acknowledges financial support from the Italian Ministry of Research and Education (MIUR) by a starting package of Politecnico di Torino, grant number 54\_RSG20GD01.

\end{document}